\documentclass[aps,prl,twocolumn,showpacs]{revtex4}

\usepackage{graphicx}

\begin{document}


\title{Impurity Scattering Effect on Superconductivity and the Violation of Anderson Theorem in Ba(Fe$_{1-x}$Co$_{x}$)$_{2}$As$_{2}$ Single Crystals}

\author{Gang Mu, Bin Zeng, Peng Cheng, Zhaosheng Wang, Lei Fang, Bing Shen, Lei Shan, Cong Ren, and Hai-Hu
Wen$^{\star}$} \affiliation{National Laboratory for
Superconductivity, Institute of Physics and Beijing National
Laboratory for Condensed Matter Physics, Chinese Academy of
Sciences, P.O. Box 603, Beijing 100190, China}

\begin{abstract}
Low-temperature specific heat (SH) and resistivity were measured on
Ba(Fe$_{1-x}$Co$_{x}$)$_{2}$As$_{2}$ single crystals in wide doping
region. A sizeable residual specific heat coefficient $\gamma_0$ was
observed in the low temperature limit of all samples. The specific
heat jump near $T_c$, i.e. $\Delta C/T|_{T_c}$ and the upper
critical field H$_{c2}^c$ (H$||$c) were also determined. It is found
that $-\gamma_0$, $\Delta C/T|_{T_c}$, $\sqrt{H_{c2}^c}$ and $T_c$
all shared a similar evolution with doping. All these can be well
understood within the model of S$^\pm$ pairing symmetry when
accounting the Co-dopants as unitary scattering centers in the FeAs
planes. Our results give a direct evidence for the violation of the
Anderson theorem in FeAs-based superconductors.
\end{abstract}

\pacs{74.20.Rp, 74.70.Dd, 74.62.Dh, 65.40.Ba} \maketitle

The discovery of superconductivity in iron pnictides have generated
enormous interests in the community of condensed matter
physics.\cite{Kamihara2008} One of the key issues here is about the
superconductivity mechanism. Phenomenologically it has been found
that in many, if not all, structures of the iron-pnictide materials,
the parent phase has a long range antiferromagnetic (AF) ordered
state,\cite{DaiPCNature} and the superconductivity is induced by
suppressing this AF order.\cite{ChenGFCeFeAsF,BaK,SrF,LaSr} Another
important issue concerning the superconductivity mechanism is about
the pairing symmetry. Up to date, experimental results gave rather
contradicting conclusions about the pairing symmetry in the iron
pnictide
superconductors.\cite{WYL,Sato,ZhGQ,MuG,Chien,HDing,Hashimoto}
Theoretically it was suggested that the superconducting pairing may
be established via exchanging spin fluctuations between the
electrons in the hole pockets (around $\Gamma$ point) and the
electron pockets (around M point),\cite{Mazin,Kuroki} thus a model
concerning s-wave symmetry with opposite sign between different
bands (the so-called S$^\pm$) was proposed. A direct evidence to
prove this unique pairing manner is still lacking although some
indirect evidence does indicate that the superconductivity vanishes
gradually when the condition for this interpocket scattering
deteriorates.\cite{HDing2,LFang}

In a conventional superconductor, the non-magnetic impurity will not
lead to apparent pair-breaking effect, therefore no quasiparticle
density of states (DOS) can be generated at the Fermi energy E$_F$.
This was called as the Anderson theorem.\cite{AndersonTheorem} In
sharp contrast, in a d-wave superconductor, non-magnetic impurities
can induce a high DOS due to the existence of nodes. Thus it is not
strange when a large residual specific heat coefficient $\gamma_0$
(which measures actually the DOS at E$_F$) was observed in the
cuprate superconductors.\cite{WenHHPRB2004} As for the case with the
pairing symmetry of S$^\pm$, it has been pointed out that
non-magnetic impurity disorder could severely suppress $T_c$ and the
gap.\cite{Cvetkovic} Recently, the effect of impurity scattering in
the case of S$^\pm$ pairing symmetry was considered and it was found
that the DOS spectrum $\rho (\omega)$ can be significantly modified
leading to a finite DOS at E$_F$.\cite{Bang,Parker,Onari,Ng}
Conclusions about the residual DOS from thermal transport
measurement are controversial with each
other.\cite{XGLuo,SYLi,Machida,Shibauchi} To verify this theoretical
hypothesis, specific heat (SH) measurement may be a good choice
because it is straightforward to get the information of DOS at the
Fermi level. In this Letter, we report the low-temperature SH under
different magnetic fields on the
Ba(Fe$_{1-x}$Co$_{x}$)$_{2}$As$_{2}$ single crystals from underdoped
to overdoped region. We found a sizeable value of $\gamma_0$ for all
samples in the low temperature limit. The doping dependence of
$\gamma_0$ anti-correlates with that of $T_c$, $\Delta C/T|_{T_c}$
and $\sqrt{H_{c2}^c}$. These behaviors were well interpreted within
the model of S$^\pm$ pairing symmetry.

\begin{figure}
\includegraphics[width=8cm]{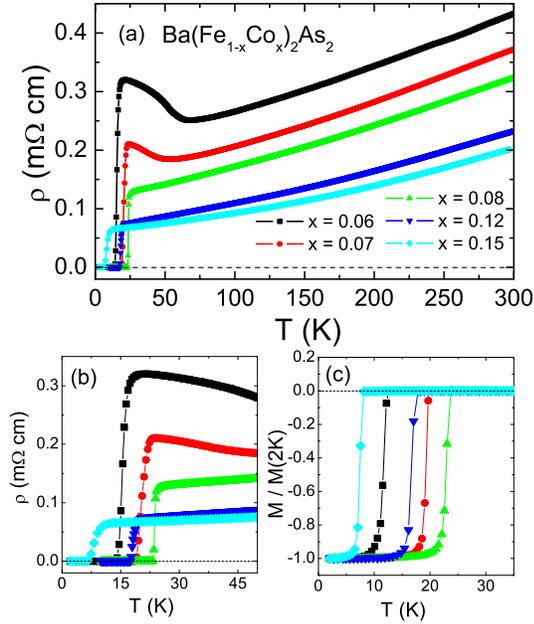}
\caption {(color online) (a) Temperature dependence of resistivity
for the Ba(Fe$_{1-x}$Co$_{x}$)$_{2}$As$_{2}$ single crystals in wide
doping range under zero field. (b)The enlarged view of the
resistivity data near the superconducting transition.  (c)The dc
magnetization data measured with $H$ = 20 Oe for the zero field
cooling (ZFC) process. The curves were normalized by the
magnetization data obtained at 2 K.  } \label{fig1}
\end{figure}

The Co-doped Ba(Fe$_{1-x}$Co$_{x}$)$_{2}$As$_{2}$ single crystals
were grown by the self-flux method\cite{LFang}. The samples for the
SH measurement have typical dimensions of 2.5 $\times$ 1.5 $\times$
0.2 mm$^{3}$. The dc magnetization measurements were done with a
superconducting quantum interference device (Quantum Design, SQUID,
MPMS7). The resistivity and the specific heat were measured with a
Quantum Design instrument physical property measurement system
(PPMS) with the temperature down to 1.8 K and the magnetic field up
to 9 T. We employed the thermal relaxation technique to perform the
specific heat measurements. To improve the resolution, we used a
latest developed SH measuring frame from Quantum Design, which has
negligible field dependence of the sensor of the thermometer on the
chip as well as the thermal conductance of the thermal linking
wires.

In Fig. 1(a), we show the temperature dependence of resistivity
under zero field for five samples with nominal doping compositions
of $x$ = 0.06, 0.07, 0.08, 0.12, and 0.15, respectively. The sample
with $x$ = 0.08 was found to be optimally doped with the highest
onset transition temperature $T_c^{onset} \approx 24.5$ K. In the
underdoped region ($x < 0.08$), an upturn in the resistivity curve
above $T_c$ can be easily seen, which was supposed to be related
with the structural and antiferromagnetic (AF)
transition.\cite{LFang} An enlarged view of the $\rho(T)$ curves
near T$_c$ was shown in Fig. 1(b). We also measured the dc
magnetization of the samples, which was displayed in Fig. 1(c). The
rather sharp transitions suggest the high quality of our samples.

We show the raw data of SH for the sample $x$ = 0.08 in the main
frame of Fig. 2. The red solid line displays roughly the tendency of
the normal state SH, $C_{norm}/T = \gamma_n + C_{ph}/T$, where
$\gamma_n$ is the electronic contribution and $C_{ph}/T$ is the
phonon contribution obtained based on a simple polynomial fit in the
normal state. This will not be relied on to analyze our data. A
clear anomaly due to the superconducting transition can be observed
at about 23 K in the zero-field data. A magnetic field of 9 T
suppresses the anomaly remarkably and also moves the transition to
lower temperatures. We also show the enlarged view of the data in
the low-T region in Fig. 2(a). One can see the roughly linear
behavior in the $C/T$ vs $T^2$ plot in low-T region. Surprisingly,
no clear Schottky anomaly was detected in all the samples, which may
suggest that the Co-doping here induces no local magnetic moment
which would on the other hand give a large contribution to SH as the
Schottky anomaly. It is clear that the magnetic field enhances the
low-T SH continuously, indicating the increase of quasiparticle DOS
at E$_F$ induced by magnetic field. We will discuss this issue
later.

\begin{figure}
\includegraphics[width=9cm]{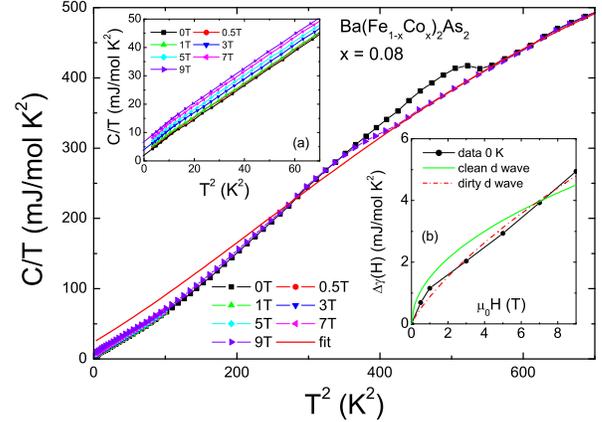}
\caption {(color online) Main frame: The raw data of SH for the
optimally doped sample ($x$ = 0.08) under different fields. The red
solid line is guide to eyes which displays the rough tendency of the
normal state SH (see text). The inset (a) shows an enlarged view of
the same data in the low-T region. Inset (b) shows the field
dependence of the field-induced term $\Delta\gamma(H) =
[C(H)-C(0)]/T$ at 0 K obtained by linearly extrapolating the low-T
data to zero temperature. The green solid, and red dash-dotted lines
are the fit to the clean d-wave prediction $\Delta\gamma(H) =
A\sqrt{H}$, and the dirty d-wave prediction $\Delta\gamma(H) =
\Lambda (H/H_{c2})\texttt{log}[B(H_{c2}/H)]$, respectively. }
\label{fig2}
\end{figure}

In order to have a comprehensive understanding, we also measured the
temperature and field dependence of SH on samples in wide doping
region. We extracted the SH difference between 0 T and 9 T and
showed the results in Fig. 3(a). From the main frame of Fig. 2, we
can see that a magnetic field of 9 T can not suppress the
superconductivity completely, but it shifts the superconducting
transition to a distinguishable lower temperature. As a result, we
can evaluate the height of the SH anomaly $\Delta C/T|_{T_c}$ near
$T_c$ from the difference of $C/T$ at 0 T and 9 T. It is clear that
the optimal doped sample with highest $T_c$ has the largest anomaly
$\Delta C/T|_{T_c}\approx$ 28.6 mJ/mol K$^2$. This value is quite
comparable with that reported by other
groups.\cite{Fisher,Canfield,Canfield2} In each doping side
(underdoping or overdoping), $\Delta C/T|_{T_c}$ seems to display a
monotonic increase with $T_c$. This behavior is qualitatively
consistent with that reported in Ref.\cite{Canfield2} where a
scaling behavior of $\Delta C/T|_{T_c} \propto (T_c)^2$ was
reported. However, we note that there is a clear difference between
the underdoped and the overdoped regions. For example, the sample
with $x$ = 0.07 has a higher $T_c$ while showing a smaller $\Delta
C/T|_{T_c}$ compared with that of the sample with $x$ = 0.12.
 As will be
discussed below, we attribute this difference in the underdoping and
overdoping regions to the different mechanism that governs the
evolvement of $\Delta C/T|_{T_c}$ with $T_c$.

\begin{figure}
\includegraphics[width=8cm]{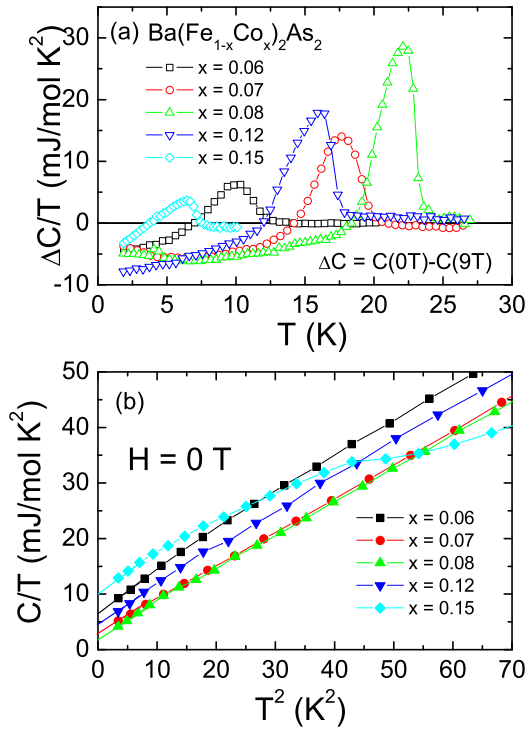}
\caption {(color online) (a) The difference of SH between 0 T and 9
T for samples in wide doping region. (b) Low-temperature SH data for
samples with different doping levels under zero temperature. One can
see a nonmonotonic evolvement of the residual value $\gamma_0$ with
doping. The departure from linear behavior at about 7 K for the
sample with $x$ = 0.15 was caused by the superconducting
transition.} \label{fig3}
\end{figure}

In Fig. 3(b), we present the low-T SH data at zero field. A linear
extrapolation of the low-T data finds immediately that there is a
sizeable value of the residual SH coefficient $\gamma_0$ for all
samples. A closer scrutiny realizes that $\gamma_0$ has a
nonmonotonic doping dependence on the Co-doping concentration $x$.
We must stress that the sizeable value of $\gamma_0$ found in
present samples should not be simply attributed to the
non-superconducting fraction. The reasons are as following: (1) A
minimum of $\gamma_0$ was observed just at the optimal doping point
(see below). From the chemistry point of view, however, there is no
reason to believe that the non-superconducting fraction should be
the lowest in the optimally doped sample; (2) Counting the
magnetization signal in the low temperature region finds that the
magnetic shielding is beyond 95 $\%$ for all the samples.

In order to clarify the origin of $\gamma_0$ in present samples, we
have extrapolated the low-T SH data shown in inset (a) of Fig. 2 to
zero temperature linearly and obtained the field-induced term
$\Delta\gamma(H) = [C(H)-C(0)]/T$ at 0 K, which was shown in inset
(b) of Fig. 2 for the optimally doped sample. One can see that
$\Delta\gamma(H)$ rises up quickly and shows a nonlinear tendency
below about 1 T. Whereas it displays a clear linear behavior above 1
T. Similar feature was observed in samples with other doping levels.
We at first attempted to fit the data with the relation
$\Delta\gamma(H) = A\sqrt{H}$ predicted for d-wave symmetry in the
clean limit.\cite{Volovik} The result was displayed by the green
solid line in the inset (b) of Fig.2. It is clear that this fitting
curve cannot describe the experimental data at all. Secondly, we
fitted our data using the relation for d-wave superconductors in the
dirty limit, $\Delta\gamma(H) = \Lambda
(H/H_{c2}^c)\texttt{log}[B(H_{c2}^c/H)]$.\cite{Hirschfeld} Here $B$
is a constant which approximates 7.26 for a triangular vortex
lattice. We left $\Lambda$ and $H_{c2}^c$ as the free fitting
parameters. The best fitting result was shown by the red dash-dotted
line. Again this curve departs from the experimental data,
especially it cannot reflect the kink feature around 1 T and the
linear feature above 1 T. So we can exclude the presence of the
superconducting gap with d-wave symmetry, either in the clean or
dirty limit. Consequently, the finite DOS found in the present
system cannot be attributed to the impurity scattering effect for a
d-wave superconductor.

The doping dependence of the extracted $\gamma_0$, along with $T_c$,
was shown in Fig. 4(a). The curve of $T_c$ vs $x$ formed an
asymmetric dome, while the $\gamma_0$ vs $x$ curve showed an
anti-correlated behavior. We argue that this behavior can be
explained by the Co-induced impurity-scattering effect. Numerical
calculations using the T-matrix method have shown that, in a
superconductor with S$^\pm$ pairing symmetry, the fully opened gap
of a clean state will be filled up by impurity states.\cite{Bang}
Therefore a finite DOS at E$_F$ may rise up (forming the so-called
gapless state) if the scattering strength becomes stronger. In the
unitary limit, the residual SH coefficient $\gamma_0$ may be
expressed by the impurity concentration $n_{imp}$ and the
superconducting gap $\Delta_s$ in a simple form $\gamma_0 \propto
(n_{imp}/\Delta_s)^\alpha$, with $\alpha>0$.\cite{Preosti} The index
$\alpha$ approximates 0.5 for a d-wave
superconductor.\cite{Hirschfeld}

In the underdoped region, assuming a proportionality between
$n_{imp}$ and the number of Co-dopant, since the magnitude of
$\Delta_s$ increases with $x$ more rapidly (see discussion later on
Fig. 4(d)) than a linear increase of $n_{imp}$, therefore $\gamma_0$
was reduced with the increase of doping, one thus qualitatively
understands that $\gamma_0$ anti-correlates with $\Delta_s \propto
T_c$. Actually the realistic case is more complicated: the
superconductivity and AF states compete with each other in the
underdoped region, and there will be less and less contributions of
DOS given by the AF state in the $T$ = 0 K approach with adding more
Co-dopants into the system. These two factors lead to the
dropping-down behavior of $\gamma_0$ versus x.

In the overdoped region, the AF order was suppressed completely.
However, as we have addressed, the Co-doping will deteriorate the
spin-fluctuations and weaken the pairing strength, resulting in the
decrease of $\Delta_s$.\cite{LFang} Meanwhile $n_{imp}$($\propto x$)
keeps rising. These two factors lead to the quick increase of
$\gamma_0$ with doping in the overdoped region. In Fig.4(b) we also
showed $-\gamma_0$ together with T$_c$. Surprisingly one can see a
quite good consistency between the doping dependence of $-\gamma_0$
and $T_c$. This good consistency is understandable because
$\gamma_0$ reflects how many spin-fluctuation-mediated scattering
channels, which are responsible for the Cooper pairing, are blocked
away by the impurities.

\begin{figure}
\includegraphics[width=9.2cm]{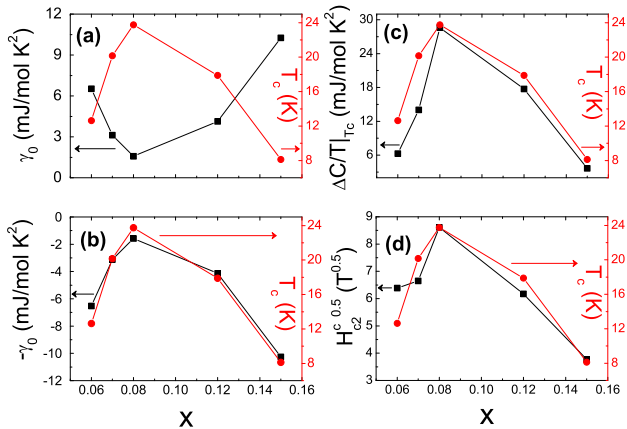}
\caption {(color online) Doping dependence of (a) the residual SH
term $\gamma_0$, (b) minus of the residual term $-\gamma_0$, (c) the
superconducting SH anomaly $\Delta C/T|_{T_c}$, and (d) square root
of the upper critical field $\sqrt{H_{c2}^c}$, plotted along with
$T_c$ for comparison.} \label{fig4}
\end{figure}

As for the doping dependence of the SH jump $\Delta C/T|_{T_c}$,
which were shown in Fig. 4(c), we can explain it based on the
variation of $\gamma_0$ with $x$. The BCS theory tells that the
height of SH jump is proportional to the effective normal state SH
coefficient $\gamma_{eff}$:
\begin{equation}
\frac{\Delta C}{T}|_{T_c} \propto \gamma_{eff}
\propto(\gamma_n^{bare} - \gamma_0)^\beta.\label{eq:1}
\end{equation}
Here $\gamma_n^{bare}$ is the bare value of SH coefficient in the
normal state (weakly dependent on doping\cite{Mazin2}) if the
impurity scattering and the competition of the AF state would not
exist, and $\beta \geq 1$. The impurity-scattering may modify the
DOS spectrum $\rho(\omega)$ in the $\omega\sim\Delta$ approach and
suppress the height of the SH jump $\Delta C/T|_{T_c}$, giving a
value of $\beta$ larger than 1. From equation (2), we can expect a
roughly consistent tendency between $\Delta C/T|_{T_c}$ and
$-\gamma_0$. Recalling the fact that $T_c$ correlates linearly with
$-\gamma_0$, one can easily see the reason for the similar
evolvement tendency between $T_c$ and $\Delta C/T|_{T_c}$ with $x$,
as shown in Fig. 4(c).

In above discussion, we have shown that the parameters $-\gamma_0$,
$\Delta C/T|_{T_c}$ and T$_c$ share a similar doping dependence. The
key player here is actually $\Delta_s$ which is estimated in the
following. In Fig. 4(d), we show the doping dependence of the square
root of the upper critical field (H$||$c-axis) $\sqrt{H_{c2}^c}$
along with $T_c$, where $H_{c2}^c$ was determined using the
Werthamer-Helfand-Hohenberg relation\cite{WHH-relation} from the
field dependent resistivity data (not shown here). One can see that
both set of data overlap quite well. This is understandable because
the Ginzburg-Landau theory has given the relation
\begin{equation}
H_{c2}^c = \frac{\Phi_0}{2\pi \xi_c\xi_{ab}} \propto
{\Delta_s}^2,\label{eq:2}
\end{equation}
where $\Phi_0$ is the flux quantum and $\xi_c$ ($\xi_{ab}$) is the
coherence length in the direction of c axis (ab plane). As a result,
$\sqrt{H_{c2}^c}$ is proportional to $\Delta_s\propto T_c$.

In summary, we studied the low-temperature SH and resistivity on
Ba(Fe$_{1-x}$Co$_{x}$)$_{2}$As$_{2}$ single crystals in wide doping
region. A sizeable residual SH coefficient $\gamma_0$ in the low-T
limit and clear SH jumps were detected in all samples. It is found
that $-\gamma_0$, $\Delta C/T|_{T_c}$, $\sqrt{H_{c2}^c}$, and $T_c$
all share a similar evolution with doping amount $x$. All these
behaviors were interpreted within the model of S$^\pm$ pairing
symmetry considering the Co-doping induced scattering effect in this
system.

\begin{acknowledgments}
We appreciate the useful discussions with Jan Zaanen and Dung-Hai
Lee. This work is supported by the NSF of China, the Ministry of
Science and Technology of China (973 projects: 2006CB601000,
2006CB921107, 2006CB921802), and Chinese Academy of Sciences
(Project ITSNEM).
\end{acknowledgments}

 $^{\star}$ hhwen@aphy.iphy.ac.cn


\begin{thebibliography}{99}

\bibitem{Kamihara2008} Y. Kamihara \emph{et al.}, J. Am. Chem. Soc. {\bf130}, 3296 (2008).
\bibitem{DaiPCNature}  Clarina de la Cruz \emph{et al.}, Nature {\bf453}, 899 (2008).
\bibitem{ChenGFCeFeAsF} G. F. Chen \emph{et al.}, Phys. Rev. Lett. {\bf100}, 247002 (2008).
\bibitem{BaK} M. Rotter, M. Tegel, and D. Johrendt, Phys. Rev. Lett. {\bf101}, 107006 (2008).
\bibitem{SrF} F. Han \emph{et al.}, Phys. Rev. B {\bf78}, 180503(R) (2008).
\bibitem{LaSr} H. H. Wen \emph{et al.}, Europhys. Lett. {\bf82}, 17009 (2008).
\bibitem{WYL} Y. Wang \emph{et al.}, Supercond. Sci. Technol. {\bf22}, 015018 (2009).
\bibitem{Sato} T. Sato \emph{et al.}, J. Phys. Soc. Jpn. {\bf77}, 063708 (2008).
\bibitem{ZhGQ} S. Kawasaki \emph{et al.}, Phys. Rev. B {\bf78}, 220506(R) (2008).
\bibitem{MuG} G. Mu \emph{et al.}, Chin. Phys. Lett. {\bf25}, 2221 (2008).
\bibitem{Chien} T. Y. Chen \emph{et al.}, Nature (London) {\bf453}, 1224 (2008).
\bibitem{HDing} H. Ding \emph{et al.}, Europhys. Lett. {\bf83}, 47001 (2008).
\bibitem{Hashimoto}K. Hashimoto \emph{et al.}, Phys. Rev. Lett. {\bf102}, 017002 (2009).
\bibitem{Mazin} I.I. Mazin \emph{et al.}, Phys. Rev. Lett. {\bf101}, 057003 (2008).
\bibitem{Kuroki} K. Kuroki \emph{et al.}, Phys. Rev. Lett. {\bf101}, 087004 (2008).
\bibitem{HDing2} Y. Sekiba \emph{et al.}, N. J. Phys. {\bf11}, 025020 (2009).
\bibitem{LFang}  L. Fang \emph{et al.}, arXiv: 0903.2418.
\bibitem{AndersonTheorem} P. W. Anderson, J. Phys. Chem. Solids {\bf11}, 26 (1959).
\bibitem{WenHHPRB2004} H. H. Wen \emph{et al.}, Phys. Rev. B {\bf70}, 214505 (2004).
\bibitem{Cvetkovic}  V. Cvetkovic, and Z. Tesanovic, Europhys. Lett. {\bf85}, 37002 (2009).
\bibitem{Bang} Y. Bang, H. Choi, and H. Won, Phys. Rev. B {\bf79}, 054529 (2009).
\bibitem{Parker} D. Parker \emph{et al.}, Phys. Rev. B {\bf78}, 134524 (2008).
\bibitem{Onari} S. Onari, and H. Kontani, arXiv: 0906.2269.
\bibitem{Ng} T.K. Ng, and Y. Avishai, arXiv: 0906.2442.
\bibitem{XGLuo}  X. G. Luo \emph{et al.}, arXiv: 0904.4049.
\bibitem{SYLi}  L. Ding \emph{et al.}, arXiv: 0906.0138.
\bibitem{Machida}  Y. Machida \emph{et al.}, arXiv: 0906.0508.
\bibitem{Shibauchi}  M. Yamashita \emph{et al.}, arXiv: 0906.0622.
\bibitem{Fisher} J. Chu \emph{et al.}, Phys. Rev. B {\bf79}, 014506 (2009).
\bibitem{Canfield} N. Ni \emph{et al.}, arXiv: 0905.4894.
\bibitem{Canfield2} S. L. Bud¡¯ko, N. Ni, and P. C. Canfield, arXiv: 0905.2955.
\bibitem{Volovik} G. E. Volovik, JETP Lett. {\bf58}, 469 (1993); {\bf65}, 491 (1997).
\bibitem{Hirschfeld} C. K\"{u}bert, and P.J. Hirschfeld, Solid State Commun. {\bf105}, 459 (1998).
\bibitem{Preosti} G. Preosti, H. Kim, and P. Muzikar, Phys. Rev. B {\bf50}, 1259 (1994).
\bibitem{Mazin2}I. Mazin, Priviate Communication.
\bibitem{WHH-relation} N. R. Werthamer, E. Helfand, and P. C. Hohenberg, Phys. Rev. {\bf147}, 295 (1966).


\end{thebibliography}
\end{document}